\documentclass[traditabstract]{aa} 
\usepackage{graphicx}
\usepackage{txfonts}
\usepackage{natbib}
\bibpunct{(}{)}{;}{a}{}{,}
\def \lya {Lyman-$\alpha\,$}
\def \rp {R$_p$}
\def \hdn {HD\,189733}
\begin{document}
   \title{Hubble Space Telescope detection of oxygen in the atmosphere of exoplanet \hdn b}
\titlerunning{Detection of oxygen in exoplanet \hdn b}
\authorrunning{Ben-Jaffel \& Ballester}

   \author{L. Ben-Jaffel
          \inst{1, 2}
          \and
          G. E. Ballester \inst{3} }

   \institute{UPMC Univ Paris 06, UMR7095, Institut d'Astrophysique de Paris,
              F-75014 Paris, France. \email{bjaffel@iap.fr}
         \and
               CNRS, UMR7095, Institut d'Astrophysique de Paris, F-75014, Paris, France
         \and
              Univ. of Arizona, Lunar \& Planetary Lab., 1541 E. Univ. Blvd,Tucson, AZ 85721-0063
             \email{gilda@lpl.arizona.edu} }

   \date{Received 28 December 2012; accepted 25 February 2013}

\abstract{Detecting heavy atoms in the inflated atmospheres of giant exoplanets that orbit close to their parent stars is a key factor for understanding their bulk composition, their evolution, and the processes that drive their expansion and interaction with the impinging stellar wind. Unfortunately, very few detections have been made thus far. Here, we use archive data obtained with the Cosmic Origins Spectrograph onboard the Hubble Space Telescope to report an absorption of $\sim 6.4\%\pm1.8\%$ by neutral oxygen during the \hdn b transit. Using published results from a simple hydrodynamic model of HD\,189733b, and assuming a mean temperature of $\sim (8-12)\times10^3$\,K for the upper atmosphere of the exoplanet, a mean vertical integrated O\,I density column of $\sim 8\times 10^{15}\,{\rm cm^{-2}}$ produces only a $3.5\%$ attenuation transit. Much like the case of the hot-Jupiter HD\,209458b,  super-solar abundances and/or super-thermal broadening of the absorption lines are required to fit the deep transit drop-off observed in most far-ultraviolet lines. We also report evidence of short-time variability in the measured stellar flux, a variability that we analyze using time series derived from the time-tagged exposures, which we then compare to solar flaring activity. In that frame, we find that  non-statistical uncertainties in the measured fluxes are not negligible, which calls for caution when reporting transit absorptions. Despite cumulative uncertainties that originate from variability in the stellar and sky background signals and in the instrument response, we also show a possible detection for both a transit and early-ingress absorption in the ion C\,II 133.5 nm lines. If confirmed, this would be the second exoplanet for which an early ingress  absorption is reported. In contrast, such an early ingress signature is not detected for neutral O\,I.  Assuming the \hdn b magnetosphere to be at the origin of the early absorption, we use the Parker model for the stellar wind and a particle-in-cell code for the magnetosphere to show that its orientation should be deflected $\sim 10-30^{\circ}$ from the planet-star line, while its nose's position should be at least $\sim 16.7$\,\rp\ upstream of the exoplanet in order to fit the C\,II transit light curve. The derived stand-off distance is consistent with a surface  magnetic field strength of $\sim 5.3$\,Gauss for the exoplanet, and a supersonic stellar wind impinging at $\sim 250$\,km/s, with a temperature of $1.2\times 10^5$\,K and a density $\sim 6.3\times 10^6\,{\rm cm^{-3}}$ at the planetary orbit, yet the  fit is not unique.}  

\keywords{planet-star interactions-- planets and satellites: magnetic fields -- stars: individual: HD\,189733-- star: activity ultraviolet -- sun: activity -- ultraviolet: general}
\maketitle
%

\section{Introduction} 

Observations  in the far-ultraviolet (FUV) spectral range of HD\,209458b, the first detected exoplanet transiting a  sun-like star, have clearly shown that its atmosphere is hot and inflated \citep{vid03,bal07,ben07}. However, there is no clear spectral indication on any escape of gas outside its Roche lobe \citep{ben08}. Filling the Roche lobe is a necessary but not sufficient condition for escape to occur since various effects could confine the extended nebula. Planetary magnetic field, even of weak strength, will trap particles on closed field lines, as shown by several studies \citep{ada11,tra11}. The impinging stellar wind can also confine the expanding gas \citep{bis13,mur09,sto09}. In addition, the planetary magnetic field could affect ions and neutrals differently \citep{kho12}. The level of stellar X-ray and extreme-ultraviolet (XUV) radiation at the planet as well as time variations can also affect the extension of the neutral versus ion components \citep{guo11}.  All these effects cast doubts on the distribution of the upper atmospheric components assumed to produce the transit absorption signatures, and thus on the mass loss rates disseminated thus far in the literature \citep[e.g.,][]{lam12}.

 Two main difficulties, the relative faintness of the FUV (115-143 nm) stellar emissions and the variability of the sources,  are at the origin of the major uncertainties on any estimation of the composition and the net mass loss from a exoplanetary atmosphere. Indeed, of the 854 exoplanets detected until the end of 2012, only 291 are transiting and very few systems are close enough to the Earth to show a high enough FUV flux level to be detected by the Hubble Space Telescope (HST) without requiring very expensive resources. This explains that up to now, only a few transiting planets have been observed in the FUV. Furthermore, the two most aften observed stars, namely the  sun-like HD\,209458 star and the young K-star \hdn, show clear evidence of variability in their flux on time scales ranging from a few hours to the rotational period \citep{ben07,lec10}. Flaring activity has also been reported for \hdn\, based on nearly simultaneous X-ray and FUV observations \citep{lec12}. A third difficulty, due to signal contamination by both the sky background 
\begin{table*}
\caption{Archival HST COS data set used in this study}
\begin{center}
\begin{tabular}{cccc}
\hline
Data set name &   Exposure time (s) &  Orbital phase & HST orbit  \\
    lb0ukq   & 209 & -0.071 & 0   \\ 
    lb0umq   & 889 & -0.050 & 1      \\
    lb0uoq   & 889 & -0.045 & 1       \\
    lbauqq   & 889 & -0.020 & 2       \\
    lbausq   & 889 & -0.015 & 2       \\
    lbauuq   & 889 & -0.009 & 2       \\
    lbav0q   & 889 & +0.010 & 3       \\
    lbav2q   & 889 & +0.015 & 3       \\
    lbav4q   & 889 & +0.021 & 3       \\
    lbavdq   & 889 & +0.040 & 4      \\
    lbavmq   & 889 & +0.045 & 4      \\
    lbavpq   & 889 & +0.051 & 4      \\
\hline
\end{tabular} 
\end{center}

\tablefoot{The twelve individual exposures were obtained with the G130M grating in the timetag mode (program HST-GO-11673). Central transit time was defined by propagating from zero phase on HJD 2453988.80339 and a period of 2.21857312 days \citep{tri09}. The short exposure \#0 has not been used in this work.}
\end{table*}
 signal (at some spectral lines) and the instrument response,  adds to the complexity of the problem. While the first difficulty can be resolved by focusing on a few close-by and UV-bright stars, the signal variability from both the source and the instrument is a real problem that should be addressed to build a reliable diagnostic to extract an accurate description of the upper atmosphere and of the interaction region between the exoplanet and the impinging wind from its parent star.  

Before one can conclude on the relevance of the key processes that shape the upper atmosphere of hot-Jupiter exoplanets,  it is important to know their basic composition. The FUV regime is suitable for detecting atoms that show resonance lines in that spectral window. For the exoplanet HD 209458b, H\,I, O\,I, and C\,II were detected in the FUV, while doubts remain on Si\,III \citep{bal13}. For both \hdn b and 55\,Cnc\,b, only H\,I has been detected so far \citep{lec10,lec12,ehr12}. Other atomic species observed to date on HD\,189733b are neutral Na\,I \citep{hui12} and K\,I \citep{pon12}, but these species are detected at the base of the thermosphere and below (below $\sim$5000\,km altitudes) since they are readily ionized by stellar UV light. 

Here, we use archival data obtained in 2009 with the Cosmic Origins Spectrograph (COS) onboard HST to study the characteristics of the \hdn\,system in more details, and address the problem of signal variability and its effect on determinating atmospheric abundances and  properties of the star-planet interaction regions of these distant objects. To analyze the observational results, we estimate the properties of the stellar wind at the planet distance and adapt an existing model to simulate the wind's interaction with a magnetized hot-Jupiter, tailored to the specific case of the HD\,189733 system.
\section{Observations and data analysis}
The medium-resolution G130M grating was used and kept at a single wavelength setting (1291 \AA) throughout the observations, made on Sep 16 (18:31 UT) -17 (01:11 UT) 2009, to best determine relative changes in the stellar signal caused by the planetary transit (see Table 1).  The co-added spectrum of the star is shown in Figure 1. The table lists the exposures, orbital phases, and the HST orbit during which each exposure was obtained.  Here, we use the time-tagging of the COS data to derive a time series of the system signal every 100 s, following the procedure described in \cite{ben07}.  This allows us to better determine the true transit absorption signature and obtain a good track of the signal variability over selected periods. Furthermore, the time series 
\begin{figure}
\centering
\hspace*{-0.4in}
 \includegraphics[scale=0.4]{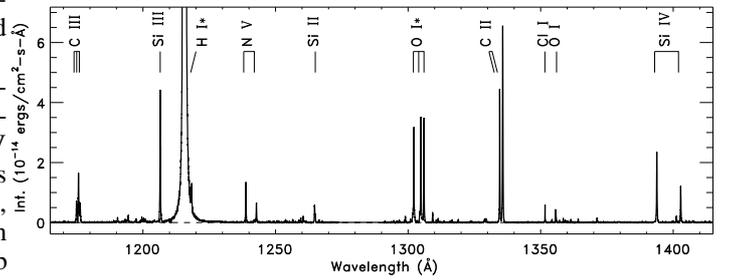}
\caption{\hdn full spectrum obtained by HST/COS G130M. The geocoronal O\,I and H\,I emissions have not been subtracted.}
\end{figure}
should reveal any sudden stellar variation that may occur during short periods. We find that although the observations did not sample the middle of the transit, they did sample a good portion of the transit. Indeed, the exoplanet's orbital phases corresponding to HST orbits 2 and 3 are inside the transit window observed for HD\,189733b in the visible \citep{tri09}. 

The derivation of the in-transit/out-of-transit flux ratios, and thus absorption by the planet disk and atmosphere, is sensitive to the choice of the out-of-transit reference exposures. Our choice here was based on the behavior of the stellar O\,I signal.  After several trials,  we opted to merge the last three exposures for the out-of-transit reference (see Table 1). This selection was also tested by using it to extract the transit signature from the moderately bright stellar lines Si\,IV 139.3 and 140.3\,nm.  Their integrated flux over the full line profiles gives an obscuration of 2.6$\pm 2.5{\%}$ (statistical), which agrees with the nominal disk obscuration of 2.5\% \citep{bou05}.

For all lines, a weak stellar continuum has been subtracted, and photon noise errors were propagated. Transit absorptions for the most relevant species are listed in Table 2. In the following, we focus on the four species, O\,I, C\,II, Si\,III, and Si\,IV for which the stellar signal is strong enough to allow a reliable study of the absorption during the planetary transit and/or to address the unavoidable variability due to all sources within the field of view and from the instrument response.  Among the selected lines, only O\,I has a background emission from the geocorona. In addition, the Si\,III 120.6\,nm emission is contaminated from the extended blue wing of the stellar Lyman-$\alpha$\,line. This is one of the reasons why we kept the Si\,III line to evaluate the time variability, but did not use it for any diagnostic of the planetary absorption. Light curves were first derived and studied statistically to evaluate the confidence we may attach to simple representations of the time series before we compared in- to out-of-transit line profiles of the key lines. The crucial problem of the time variability of the observed flux, particularly that related to the patchiness on the stellar disk is discussed in section 3.

\subsection{Oxygen triplet 130.4nm }

The 100-s time-tag sampling of the exposures does not show any steep change in the total flux for the oxygen emission in any of the triplet lines beyond statistical fluctuations as a function of time.  Significant geocoronal background emission is seen particularly in the first exposure of the HST orbit, and we successfully removed this background with a method that uses a deep-sky exposure taken by the COS team in the same grating wavelength 
\begin{figure}
\centering
 \includegraphics[width=8.cm, angle=90]{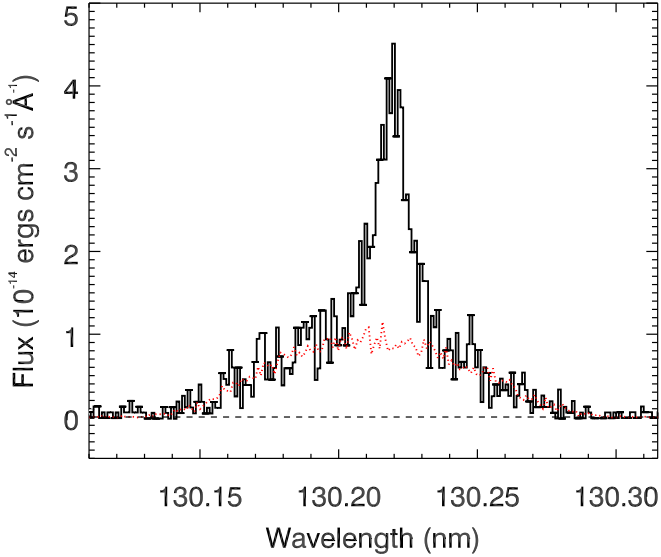}
\caption{HST/COS G130M O\,I 130.2\,nm . Solid line: combined \hdn\,and background geocoronal lines obtained on egress (dataset lbavdq). The thin feature is the star emission and the broad one is the Earth line emission filling the 2.5 arsec aperture. Dotted line: independent spectrum of the geocoronal O\,I emission line that was scaled to subtract the sky background contamination from the stellar observation.}
\end{figure}
setting (http://www.stsci.edu/hst/cos/calibration/airglow.html). First, we notice that the \hdn\,O\,I lines appear as thin and bright features over the broader gecoronal lines (Fig. 2). The additional error from the imprecision on the geocoronal subtraction represents few to 30\% of the total error depending on the emission level of the contamination. Errors shown here do include that additional uncertainty. It is important here to stress that for the broader stellar \lya\ emission, the line width is comparable to the width of the geocoronal line that fills the science aperture, as seen by COS, which prevents separating the stellar line from the background \citep{lin10}. The technique used here offers a unique opportunity to extend the use of COS to derive O\,I abundances from transiting exoplanets for which the O\,I lines are thin and bright enough to allow a spectral separation from the extended geocoronal line.

The OI 130.4\,nm feature is a triplet with lines at 130.2, 130.49, and 130.6\,nm. However, the 130.49\,nm line is contaminated on the blue wing by the stellar Si\,II 130.44\,nm line, which hinders  subtracting the geocoronal O\,I line. Figure 3 shows the transit flux ratios derived by merging the O\,I triplet lines, integrating 
\begin{figure}
\centering
 \includegraphics[width=9.cm]{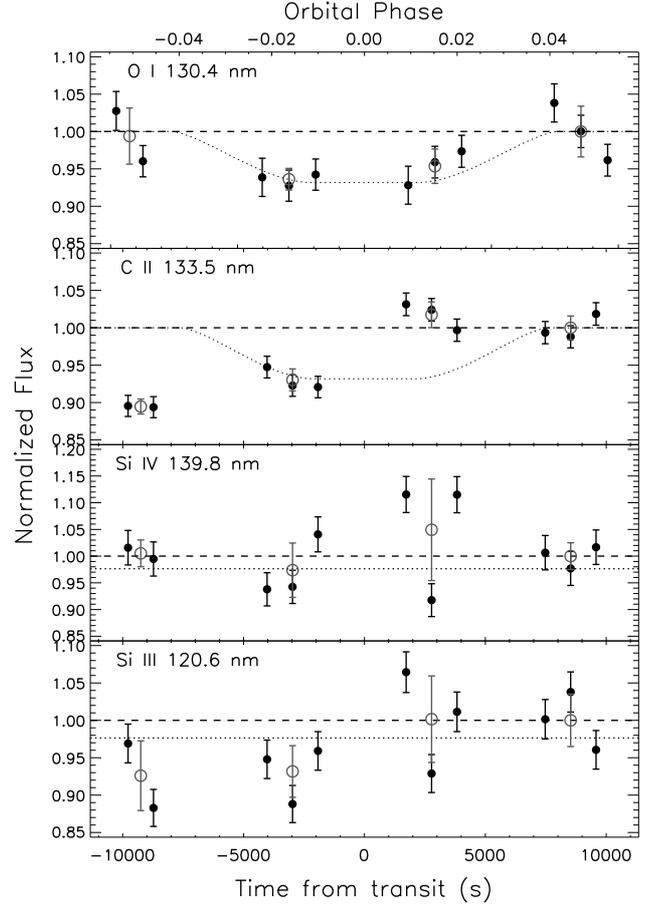}
\caption{HD\, 189733b transit light curves obtained for selected spectral lines versus time (or orbital phase): O\,I (130.4\,nm) triplet, C\,II (133.5\,nm) doublet, Si\,III (120.6\,nm), and Si\,IV (139.8\,nm) lines. Usually, those lines are used as a diagnostic of solar flux variability in the FUV \citep{bre96}. A synthetic light curve corresponding to an occulting disk of $\sim 1.7$\,\rp\,is shown with C\,II and O\,I data points. Dark error bars are statistical while the gray bars include the non-statistical scatter of the stellar signal within the given HST orbit.}
\end{figure}
each line out to $\sim\pm40 $\,km/s, since farther out there is significant error from the subtracted geocoronal background. A chi-square analysis of the OI transit light curve shows that its representation by a linear trend, which could be related to  a patchiness effect or a temporal variability in the stellar emission or a combination of these, is rejected with a confidence level slightly higher than 99\% (10 degrees of freedom). This result was obtained using the PAN package that implements the Levenberg-Marquardt technique and a bootstrap Monte Carlo simulation of the statistical noise propagation \citep{dim05}. In contrast,  assuming a synthetic light curve model (occulting disk with the radius and the time center as free parameters) gives a $\chi^2 \sim 12$ (10 degrees of freedom), a good indication that the O\,I dataset is consistent with a transit by an occulting disk of $\sim 1.7$\,\rp\,radius. This result is also confirmed by the comparison of the in-transit (orbit 2) to out-of-transit (orbit 4) line profiles for the OI 130.6\,nm line (Figure 4a). The absorption derived for the 130.2 and 130.6\,nm lines combined is $\sim8.1\pm2.2\%$, while that for the three lines is $\sim6.4\pm1.8$\%. We adopted the $\sim6.4\pm1.8\%$ absorption as the proper measurement. Despite the noise level and the limited accuracy on COS/G130M wavelengths, a transit absorption line width of $\sim 50\pm10$\,km/s is estimated for both the 130.2 and 130.6 nm lines after correcting for the line spread function (LSF). 

It is important to stress that the reported uncertainties include photon noise but no stellar/detector variability (see Table 2 for more details). The size $\sim 1.7$\,\rp\,of the O\,I opaque region falls  inside the 2.84\,\rp\, Roche lobe radius (which would produce 19$\%$ obscuration), yet this does not mean that the Roche lobe is not filled. Indeed, the intense X-ray and EUV flux from the young and active K1V star may easily ionize the gas inside the external layer between 1.7\rp\,and 2.84\rp\,that faces the star \citep{guo11}. More details on the ionized components are provided in the following section when we analyze the singly-ionized carbon lines.  Interestingly, the low-resolution detection of extended H\,I on this planet has shown a confined $\sim 5.0\pm 0.8 \%$ absorption 
\begin{figure}
\centering
 \includegraphics[width=7.2cm, angle=90]{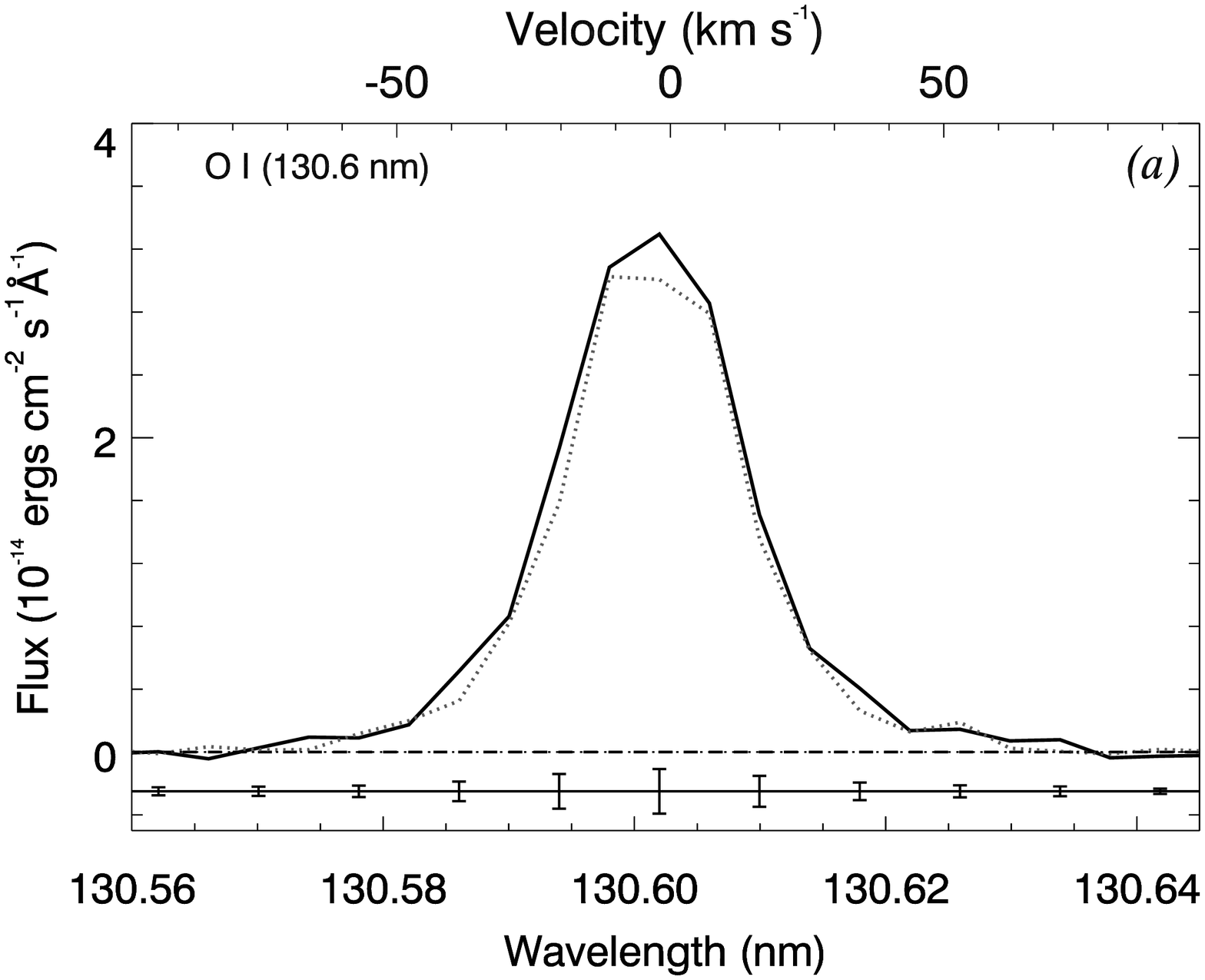}
 \includegraphics[width=7.2cm, angle=90]{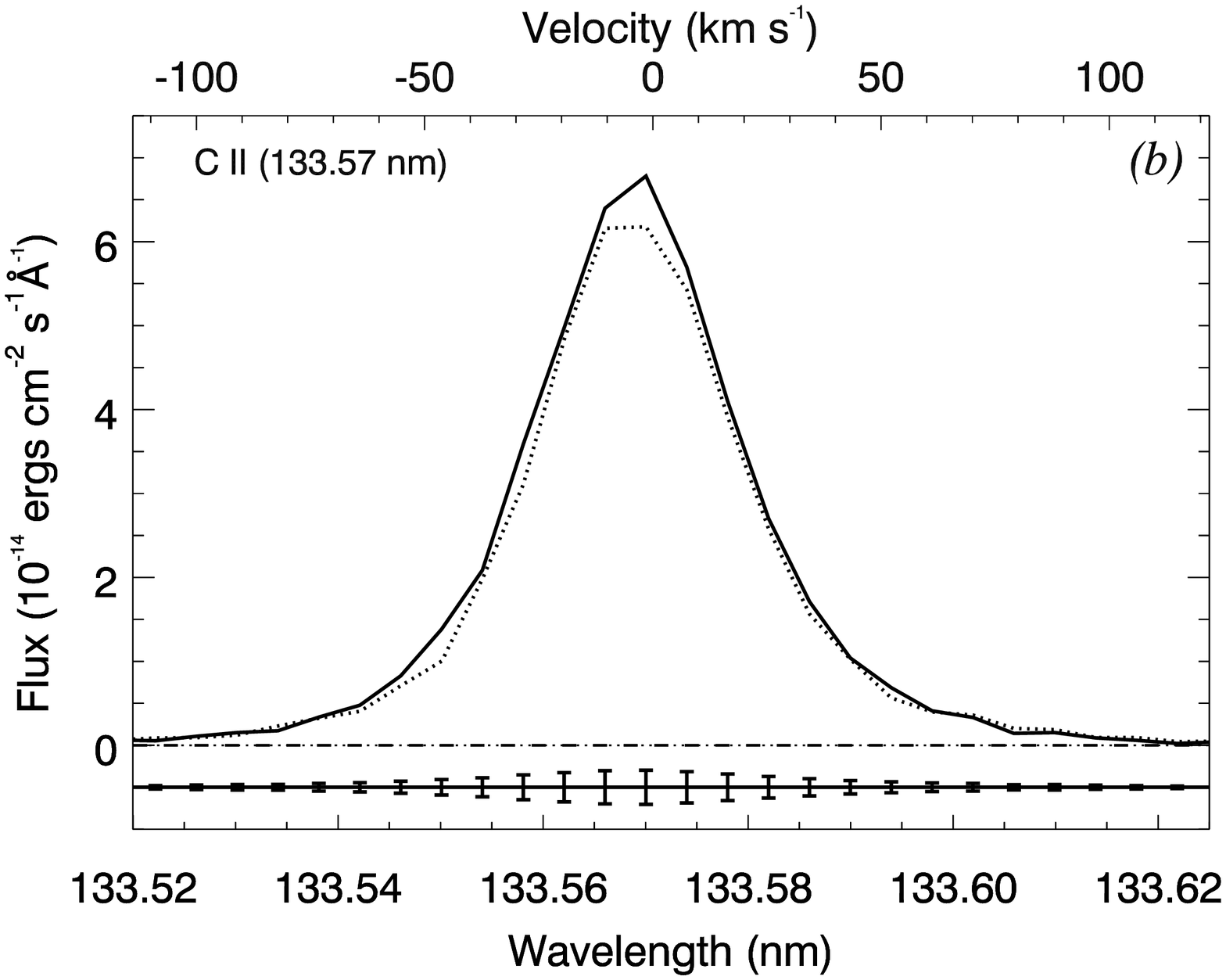}
 \includegraphics[width=7.2cm, angle=90]{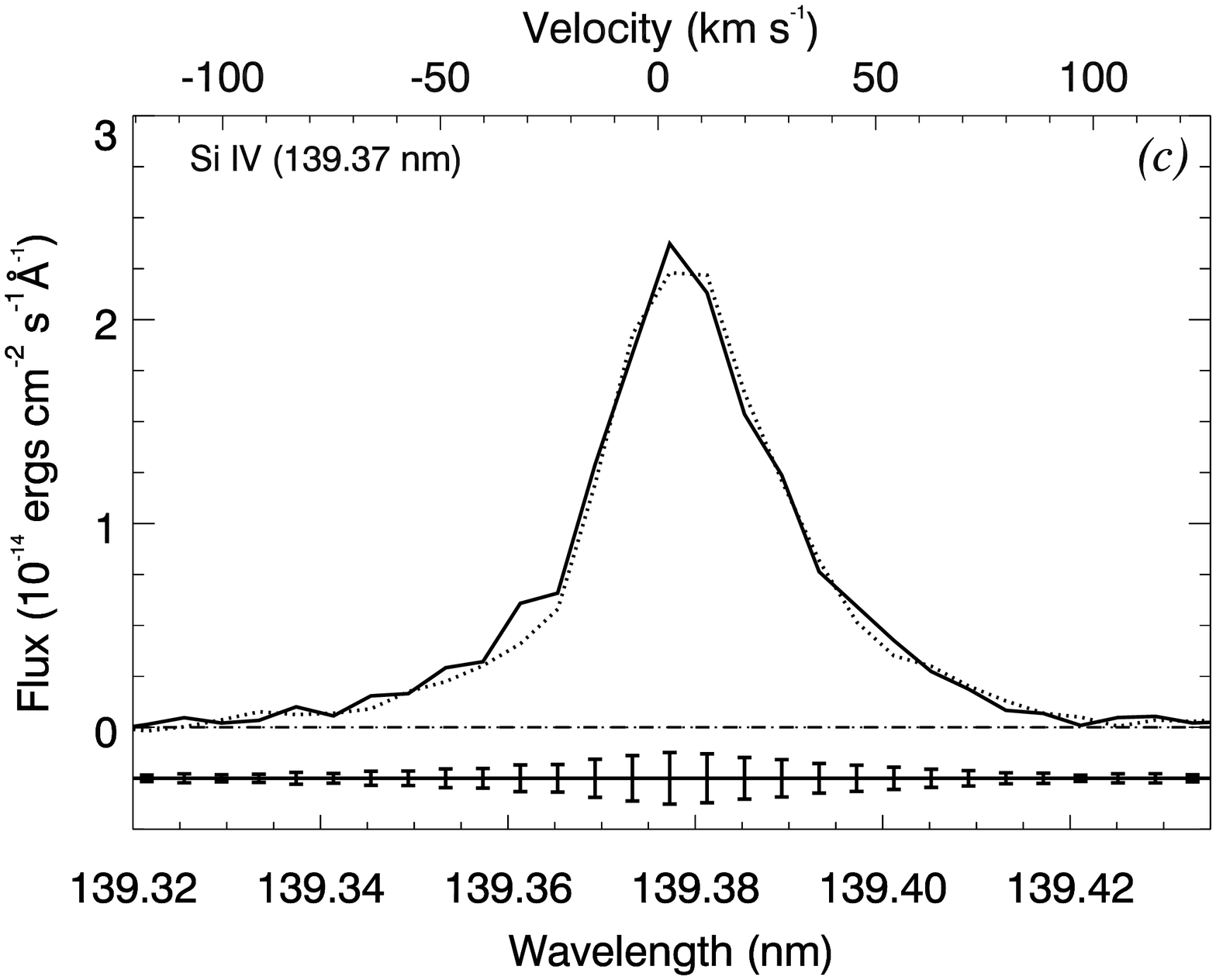}
\caption{Comparison of in-transit (orbit 2 : dotted) to out-of-transit (orbit 4: solid) line profiles. We binned spectra by 4 ($\sim 7.5$\,km/s), approximately corresponding to the nominal spectral accuracy of the dispersion of the COS/G130M grating. A redshift of $\sim 2.6$\,km/s of the star was corrected for \citep{bou05}. It is important to stress that the COS wavelength calibration does not allow one to know the relative position of the lines with an accuracy better than $\sim7.5$\,km/s \citep{oli10}. For the available low S/N COS dataset, this inaccuracy makes the interpretation of the transit absorption line profile very uncertain, particularly the veracity of any blueshift or redshift that may appear between in-transit and out-of-transit line profiles. (a) OI 130.6\,nm line. (b) C\,II 133.5\,nm line. (c) Si\,IV 139.3\,nm line.}
\end{figure}
averaged over three transits observed in 2007-2008 \citep{lec10}. Recent medium-resolution HST/STIS H\,I \lya\   transit data seems to indicate a variation with values of $2.9\pm1.4\%$ in 2010 and $5.0\pm1.3\% $ in 2011 for the line-integrated absorption \citep{lec12}. Our transit absorption obtained for neutral oxygen is thus consistent with most neutral hydrogen detections, despite the reported variations. The time variation problem for the COS data set is discussed in more detail in Sect. 3.

\subsection{Carbon doublet 133.5 nm}

This stellar emission is the brightest FUV emission after H\,I \lya\ (and about as intense as the H\,I \lya\ observed from HD\,209458), and thus provides relatively high signal-to-noise ({\it S/N}) data.  After subtracting a weak stellar continuum level, the ratios of the stellar flux per exposure versus the nominal out-of-transit data were derived by merging the C\,II 133.4 and 133.6\,nm lines, integrating each line out to $\sim \pm 50 $\,km/s from line center. These ratios are shown in Figure 3. A chi-square analysis of the C\,II time series  leads to a $\chi^2\sim40.0$ when using a linear trend (such as related to patchiness and/or temporal variation; 10 degrees of freedom), while for a representation  by two synthetic light curves, we obtain a $\chi^2\sim 28.8$ (3 free parameters: depth of planet's transit, depth and relative phase position of an extra absorption such as a magnetosphere; 9 degrees of freedom). The better confidence obtained for the two-light-curve  models may not be real because the extra freedom allowed by the model may enhance the probability that a false-positive result would be obtained. Nevertheless, the statistical analysis supports the conclusion that both the linear trend and the two-light-curve models poorly represent the data set. Obviously, a new approach that goes beyond the pure statistical analysis is required to improve our understanding of the C\,II timeseries. 

In a first step, we compared the C\,II 133.5\,nm line profiles respectively from the in-transit and ingress orbital phases to the out-of-transit line profile. As shown in Figure 4b, when compared to orbit 4, the orbit 2 (in-transit) line profile has a clear absorption that is higher than the statistical noise. Based on this C\,II line profile diagnostic, a transit absorption of $\sim 7.0 \pm 2.1\%$ is derived. Despite the noise level and the limited accuracy on COS wavelengths, we can also estimate that the spectral width of the transit absorption line is $\sim 60$\,km/s after correcting for the instrument's LSF. Similar to the transit absorption obtained for O\,I, the C\,II absorption would correspond to an opaque occulting disk of $\sim1.7$\,\rp\,that is smaller than the Roche lobe. The C\,II fluxes also show a significant absorption, at the $\sim 10\%$ level, earlier in phase (during orbit 1) than expected from ions confined to the Roche lobe (e.g., Fig 3).  The early-ingress absorption is further confirmed by the line profile diagnostic (e.g., Figure 5). This result suggests properties in the C\,II ions that are strikingly different in the O\,I neutrals in the extended upper atmosphere and beyond, where plasma interactions are at play. It is unlikely that the apparent extra absorption is caused by intrinsic stellar variation since no other stellar emission shows this temporal behavior at high {\it S/N}. The temporal variation problem is discussed in more detail in the following section to check the veracity of the observed event.  If confirmed, this would be the second target after WASP-12b for which an early-ingress absorption is detected \citep{fos10}. 

In the following, we therefore address the time variability problem of the measured fluxes, focusing on the different sources that may produce it. To distinguish the different effects, we propose a multi-species (spectral) analysis that should clarify most uncertainties and confirm our diagnostic about the O\,I detection and the potential C\,II detection. 
\begin{table*}
\caption{Absorption depth of stellar emission during the \hdn b transit} 
\begin{center}
\begin{tabular}{lcccc}
Atoms & Wavelength (nm) & Flux drop-off $^{a}$ (\%) & Photon-noise error (\%) & Total error bar (\%) \\
O\,I & $130.21+130.60$     &  $ 8.1 $ &  2.2 & 3.4\\
O\,I & $130.4$ triplet     &  $ 6.4 $  &     1.8  & 3.4  \\
C\,II & $ 133.45+133.57$     & $7.0$  & 1.2 & 2.1  \\
Si\,III  &   120.6    &   $6.8$ & 2.0 & 4.7          \\
Si\,IV   &  139.3 +140. 3   &   $2.6$  & 2.5 & 5.6             \\
\end{tabular} 
\end{center}
\tablefoot{Flux drop-off during transit of \hdn\,derived from HST/COS G130M medium-resolution observations. Total error bars include both photon noise and stellar/instrument variability.  (a) The signal drop-off is derived from the data at planet orbital phase of $\sim 0.985$ sampled during transit in HST orbit 2.}.

\end{table*}

\section{Temporal variation: a combined statistical and multi-spectral approach}

Addressing the temporal variation problem during a planetary transit is very complex because the observed variability could originate from the full stellar disk, the stellar latitudes sampled during transit, the details of the planet-obscuring area, and the instrument. Each of these sources of variability represents a challenge in itself, particularly if the data set is limited in time, which offers no opportunity to check the repeatability of any event. In the following, we discuss the three sources of variability as they may appear in the COS dataset, focusing on emissions of the four species O\,I, C\,II, Si\,III and Si\,IV. Our goal is to obtain a higher confidence in the OI and CII detections, particularly by checking if a stellar variation might mimic a transit-like light curve in the current COS dataset.

\subsection{Stellar intrinsic activity}

Stellar activity is a crucial problem in characterizating extrasolar planets and their evolution. The activity of HD\,189733 in the X-ray and EUV has been reported from XMM-Newton X-ray spectra  and from modeling of the EUV emission that was based on previous studies of stellar coronal models by \cite{for11}.  The luminosities are  log\,$L_X$ = 28.18 and  log\,$L_{EUV}$ = 28.48 in the X-ray (0.5--10\,nm) and EUV (10--92\,nm). These values are almost identical to those of the extensively studied star Epsilon Eridanis (log\,$L_X$ = 28.20 and  log\,$L_{EUV}$ = 28.44). Epsilon Eridanis and HD\,189733 are both early-K main-sequence stars (K1-K2 and K0-K1) of relatively high activity and similar estimated ages (respectively 1.1 and 1.2\,Gyr; \cite{for11}). X-ray measurements of HD\,189733 have also been made with the SWIFT satellite spanning about 30 hours in 2011, during which time enhanced activity was seen, including a bright flare \citep{lec12}.  The average of the combined X-ray and EUV fluxes derived from the SWIFT data was 7.1$\times 10^{28}$\,ergs/s. This is somewhat higher than the  total X-ray and EUV 4.5$\times10^{28}$\,ergs/s fluxes previously reported \citep{for11}.  The chromospheric Ca\,II H \& K line emission of HD\,189733 has also been measured at log$(R_{\mathrm{HK}})$=$-4.501$ with Keck \citep{knu10}. The activity of HD\,189733 is higher than that of the Sun. The mean Sun has log$L_X =$ 27.35\, erg/s  and log$(R_{\mathrm{HK}})$=$-4.905$ \citep{mam08,mam12,jud03}. For comparison, HD\,209458 may be of medium,  sun-like activity, although there are significant uncertainties \citep{bal13,for11,knu10}. 

Available FUV observations of HD\,189733 are very limited in time and offer no full reference on the activity level versus short and long time scales from existing data. What remains are the solar activity databases as a reference for the relative variation of the key FUV resonance emission lines. Our goal is to derive a reliable diagnostic on the activity level that occurs during the observing period reported in this study, using the solar flaring activity as a reference. For this purpose, we revisited Solar Radiation and Climate Experiment (SORCE) FUV archive observations of the Sun obtained since 2003 along with a catalog of flares observed by the Solar EUV Experiment on the Thermosphere Ionosphere Mesosphere Energetics and Dynamics mission \citep{woo06}.

Without any loss of generality, we may classify the stellar activity into long-term (longer than spin period), the mid-term (close to a spin period), and short-term (a day or less). Furthermore, in addition to the solar periodicity related to both the 27-day rotation and the solar cycle, flaring activity may occur at a level that can be, for strong events, comparable to the long-term activity \citep{bre96, woo03}. The variation in Si\,III, Si\,IV, H\,I, O\,I, and C\,II FUV lines have been reported for sunspot activity \citep{woo02,woo06,sno10}. However, for flare  activity, only very strong or moderate events have been published that we are aware of \citep{bre96,woo03}. To obtain several levels of activity, namely low, moderate, strong, and extreme levels of flare activity, we analyzed SORCE data obtained between 2003 and 2007, a period for which a flare activity catalog exists. Solar spectra obtained with the  Solar Stellar Irradiance Experiment (SOLSTICE) onboard SORCE on a daily basis have a low 1 nm resolution, yet they offer the opportunity to study most of our resonance lines simultaneously. After subtracting a weak solar continuum, we derived a timeseries of the irradiance of each line during the selected period. Using the list of flare events provided in the LASP database (lasp.colorado.edu/see/see-flare-catalog.html), we derived for each line an approximate estimate of the activity level that is calculated as a standard deviation around each event described in the LASP catalog (see Table 3). It is interesting to find that the variation levels are comparable for all lines for the low-activity sector, which may include variation of the plages and enhanced networks. Results differ significantly for stronger flare events. For extreme level flares (X17), the stellar variation is strong but remains comparable to the sunspot activity (see Table 3).

Here, we emphasize that the solar activity derived was used as a reference level for comparison with the \hdn\,  activity for the same resonance lines. We did not use these numbers to derive any intrinsic activity level of the HD\, 189733 FUV line emissions. We instead preferred to rely on the HST/COS data thus far obtained in the time-tag mode to derive a time series that was used to estimate stellar/instrument temporal variations that were then added quadratically to the statistical noise. In this way, the shown light curves have more realistic error bars.
\begin{figure}
\centering
 \includegraphics[width=7.2cm, angle=90]{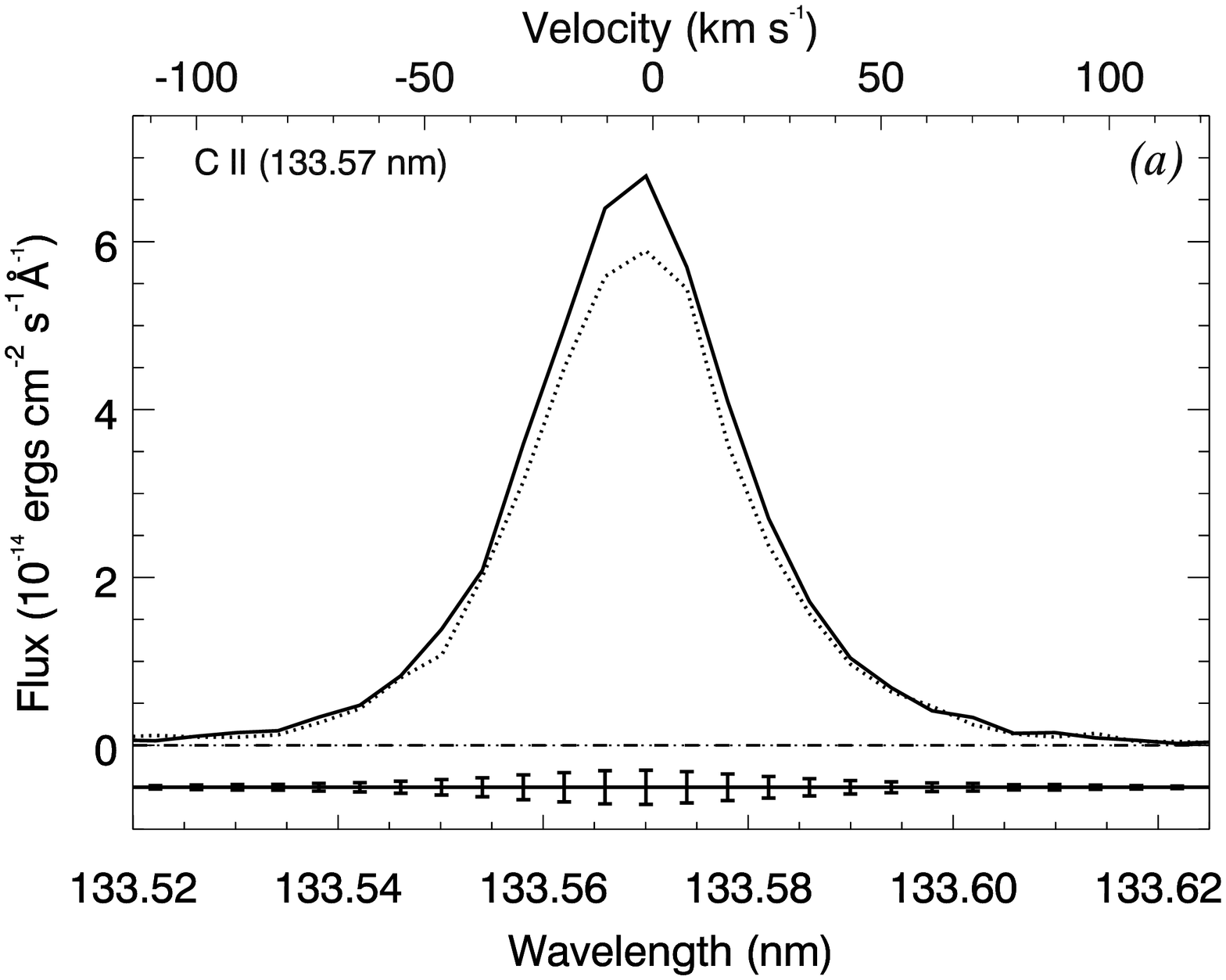}
\includegraphics[width=7.2cm, angle=90]{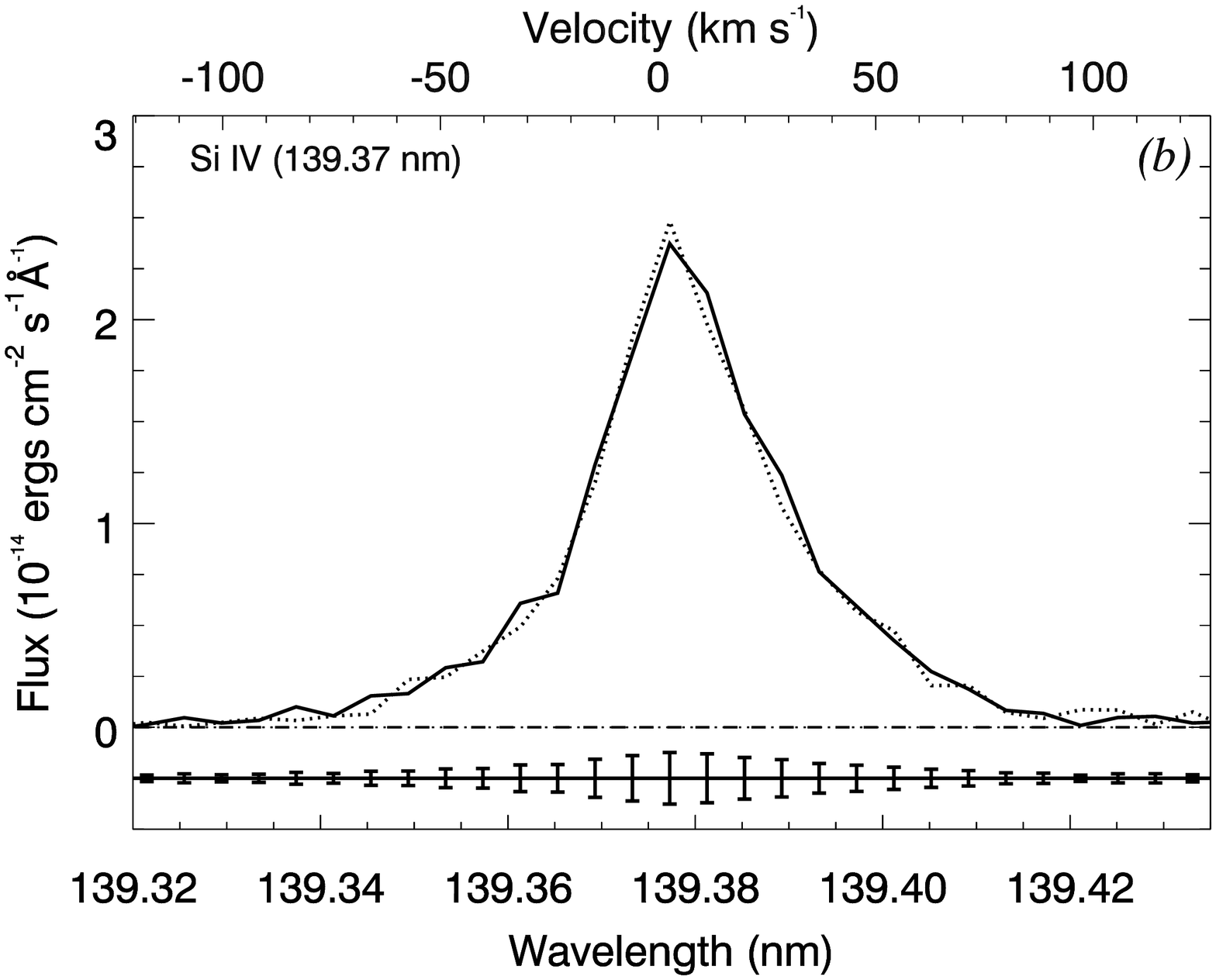}
\caption{ Comparison of ingress (orbit 1: dotted) to out-of-transit (orbit 4: solid) line profiles. We binned spectra by 4 ($\sim 7.5$\,km/s) same as in Fig. 4.  (a) C\,II 133.5\,nm line. (b) Si\,IV 139.3\,nm line.}
\end{figure}

\subsection{FUV \hdn\,flux variability: assessment}

To track stellar activity during the transit observations of HD\,189733b, we generated a time series of observed spectra every 100s. The time series has several gaps, yet it is useful for deriving key information versus the HST orbit for the spectral line selected. To this end, we first extracted the time evolution of the merging of the O\,I 130.4\,nm triplet lines, the merging of the C\,II 133.5 nm doublet, the merging of the Si\,IV 139.8\,nm doublet, and the Si\,III 120.6\,nm line (Fig. 3). In a second step, for each HST orbit and for each species, we gathered data every $\sim900$ s and derived the corresponding scatter observed between the resulting spectra during that orbit. The signal scatter was estimated as a standard deviation from the mean flux of the desired resonance line at the selected HST orbit (or transit phase angle). Finally, the photon noise level attached to the mean flux obtained during the HST orbit was subtracted quadratically from that specific standard deviation between the $\sim 900$s sub-spectra. The resulting variability in the signal level could be of stellar, planetary, or instrumental origin. To facilitate the interpretation of this variability, we plotted that extra variability versus the HST orbit for the four resonance lines selected in Figure 6. In the same figure, we also show the Sun's medium flaring activity previously derived in Table 3 as a reference. We note that a variability zero means that the signal is dominated by the photon noise. Representative photon-noise error bars are shown in Figure 6 attached to the solar value corresponding to each line.

First, we notice that most temporal variations (beyond photon noise) recorded during the total duration of the observations ($\sim 5$ hours) are weaker than the Sun's medium flaring activity. Only for orbit 3, the extra variability recorded for the Si\,IV 139.8\,nm lines is comparable to the Sun's medium flaring activity. For that orbit, the 900\,s data scatter seems to indicate a relatively strong temporal variability that is clearly visible for the Si\,III 120.6\,nm and  Si\,IV 139.8\,nm lines.  However, for the O\,I \& C\,II lines, the derived scatter of the $\sim900$\,s signal seems smaller for orbit 3 than for the Sun medium flaring activity, yet it is higher than the photon noise level. This specific behavior of the orbit 3 signal will be very useful in the following for discussing the different sources of the observed fluctuations. The similitude between signal scatter levels observed for the O\,I and C\,II lines shows that the origin is probably not the residuals from the correction of the sky background temporal variation because the C\,II lines are not affected by that contamination. For a total exposure time of $\sim 45$ minutes, we are also at a loss to explain the high variance by the response of the COS detector at the Si\,IV spectral position. In addition, as we discuss below, for the O\,I and C\,II lines, the crossing of a bright local region by the planetary obscuring area does not seem to show strong fluctuations during the 45 minutes of observation during orbit 2. This is demonstrated in figure 4, where we see for orbit 2 (during transit) comparable or lower fluctuation than in orbit 4 (out-of-transit reference orbit) for the O\,I and C\,II lines. For all these reasons, we can conclude that the signal variations observed for OI, CII, and SI IV on the orbit-to-orbit time scale are most likely of stellar and/or planetary origin (e.g., Fig. 6). The key question is how to distinguish between the two fluctuation sources on the basis of the limited COS dataset?

To begin answering this question, we stress that chromospheric OI and CII emissions are not restricted to the active regions only (the so-called patches). There are background-quiet regions that do emit in the FUV lines that originate from the chromosphere but are not that dependent on solar cycle \citep{woo00,wor01}. Since we do not know the quiet versus patchy distribution on our particular star, we cannot estimate the exact effect a particular bright patch crossing would have. However, we do know that the patchiness on the star is more pronounced for these emission lines originating from the hotter regions. Taking the Sun as a reference, we remark that the Si\,IV 139.7\,nm doublet originates from a hotter layer at log(T)$\sim4.75$ than O\,I 130.4\,nm (log(T)$\sim3.85$) and C\,II 133.5\,nm (log(T)$\sim 4.1$) \citep{woo00}. First, we need to verify whether what is known for the Sun is consistent with the HD\,189733 COS observations. Recalling  Fig. 3, we see that the O\,I and C\,II lines have much less scatter within orbits than the Si\,IV lines if the entire intra-orbit variation in orbit 3 and 4 were generated from patchiness on the star. This result is quite consistent with FUV observations of the Sun where the contrast in the plages and enhanced network regions is stronger for the Si\,IV 139.3 nm and Si\,III 120.6\,nm lines than for the O\,I 130.4\,nm and C\,II 133.5\,nm emissions \citep{wor01}. This result is very important because it allows us to use the comparison between O\,I \& C\,II from one side and Si\,IV lines from the other side to now investigate patchiness/transit effects from one orbit to another. 

\subsection{Temporal variation: final diagnostic}

For the O\,I data, our chi-square analysis rejected a linear trend with a very high confidence. In addition, the same statistical 
\begin{table*}
\caption{Solar flare activity in the FUV}
\begin{center}
\begin{tabular}{ccccc}
\hline
Species (feature nm) & Activity level &  Nature & Variability level$^a$ (\%)  \\
C\,II (133.5)     &  (L, M, H, E)  & flare & (3.7, 7.6, 12.5, 30.0)  \\ 
O\,I (130.5)     &  (L, M, H, E)  & flare & (2.7, 5.0, 7.3, 16.0)   \\ 
 Si\,III (120.6)     &  (L, M, H, E)  & flare & (4.0, 9.0, 12.8, 27.6)  \\ 
 Si\,IV (139.8)     &  (L, M, H, E)  & flare & (4.3, 8.6, 14.3, 33.4) \\ 
 C IV (154.0)     &  (L, M, H, E)  & flare & (3.9, 5.0, 11.7, 31.7)   \\ 
C\,II (133.5)     &  (27D, 11Y)  & sunspot & (22, 57)  \\ 
O\,I (130.5)     &  (27D, 11Y)   & sunspot & (15, 29)   \\ 
 Si\,III (120.6)     &  (27D, 11Y)  & sunspot & (27, 73)  \\ 
 Si\,IV (139.3)     &  (27D, 11Y)  & sunspot & (22, 60) \\ 
\hline
\end{tabular} 
\end{center}

\tablefoot{Solar observations were obtained with the Solstice instrument onboard SORCE using the 1 nm low-resolution spectrometer. The flare activity level is based on a list of flaring events derived by the Colorado/LASP team (lasp.colorado.edu/see/see-flare-catalog.html). (L, M, H, E) are for low- (class M1), medium- (class M5), high- (X1), and extreme- (X17) activity flare activity \citep{hud11}. (27D, 11Y) are the 27-days spin and solar cycle periods \citep{woo02}. (a) The variability level is estimated as a standard deviation around the selected flare event. For the 11Y period, the variability is the ratio of the mean solar irradiance value when the solar cycle is at its maximum (1992) to its minimum (1996).}
\end{table*}
\begin{figure}
\centering
 \includegraphics[width=9.cm]{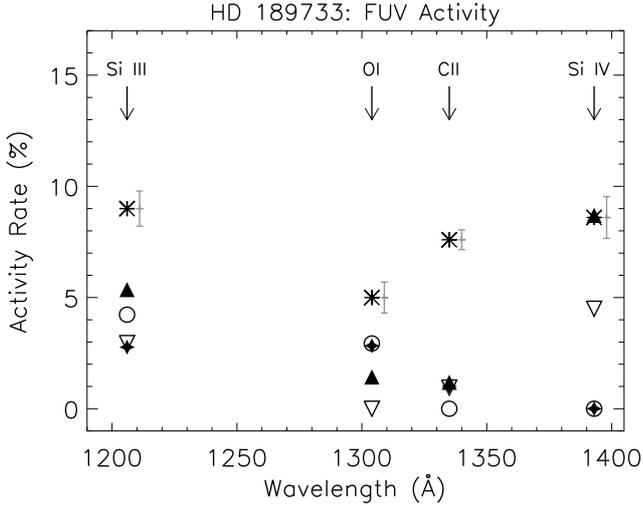}
\caption{\hdn\, activity at key species emission lines versus HST orbits during the 5-hour duration of the HST/COS G130M program used in this study. The solar flare activity is taken from Table 3. Open circles denote HST orbit 1, inverted open triangles orbit 2, filled triangles orbit 3, diamonds  orbit 4, and stars the solar medium flare activity. The corresponding statistical noise levels are shown as gray error bars.}
\end{figure}
analysis showed that a transit light curve (with the same degrees of freedom as the linear trend) was a good representation of the data. Furthermore, the comparison between in- to out-of-transit line profiles for the O\,I lines clearly shows the expected absorption. Finally, the observed flux drop-off in the O\,I signal that appears unchanged between  orbit 2 and 3 is unlikely to be produced by the planet crossing a concentration of active regions because we see no similar effect in the Si\,IV lines which are more sensitive to these regions. Indeed, we emphasize that for the Si\,IV data set, the orbit 2 average signal does show the expected transit absorption, while orbit 3 shows the opposite effect from that observed in the O\,I signal. We can therefore  conclude that the O\,I transit is detected with a high confidence.

For the C\,II dataset, assuming a linear trend is a good representation of the time variation of the signal (despite the poor confidence derived from the statistical analysis), one would expect a more pronounced effect for Si\,IV emission because the latter originates from hotter regions. As shown in Fig. 3, the orbit-to-orbit variation in the Si\,IV averaged-signal is much weaker than the one observed for C\,II. In addition, the flux averaged in orbit 2 shows a variation that is exactly what is expected from a transit for the Si\,IV line. For these reasons, we dismiss the idea that any spatial concentration of active regions on the stellar disk may have produced the C\,II light curve. The fact that the scatter is larger for Si\,IV lines during orbits 2 and 3 is probably related to bright-limb-crossing effects. Indeed, the expected limb brightening for Si\,IV (a center-to-limb factor (CTL) $\sim3.82$ increase in the intensity for the Sun) is much higher than for the O\,I and C\,II lines (CTL$\sim1.21$ for C\,II, CTL$\sim1.19$ for O\,I) \citep{wor01}. For Si\,III the bright-limb-crossing effect is also significant with CTL$\sim1.74$.  In addition, the timing of the two orbits (2 and 3) is close to the planetary disk crossing of the stellar bright limb. Therefore, the scatter thus far observed (high, low, high flux) for Si\,IV during orbit 3 is probably related to the bright-limb-crossing with the middle point being exactly the crossing time on egress. The bright-limb-crossing effect also seems to be happening in orbit 2. 

The averaged signal from orbit 3 shows for C\,II a specific behavior that is observed in all lines except O\,I, and that is not easy to  explain with simple changes in the planet absorption or by the stellar patchiness. The origin of this enhancement could be a flare event, a scenario that cannot be assessed from the limited dataset however. Despite the uncertainty on the origin of the orbit 3 signal, the C\,II line profile information from orbits 2 and 4 and the inconsistency of the patchiness assumption discussed above (see Figure 5a) both support a tentative detection of the C\,II transit. In addition, because we do not see any similar behavior in the Si\,IV lines that would result from a patchiness effect, we now have more confidence in the early-ingress absorption possibility for C\,II, particularly in view of the line profile comparison in Figure 5. Nevertheless, in all cases, we must stress that results for the C\,II data set require confirmation with future observations. The tentative detection of an early-ingress ion absorption on this planet has motivated us to model the interaction of the planet with the magnetized stellar wind. This modeling effort is a valuable tool for planning future observations of the system.

\section{Comparison with model simulations}

The O\,I lines show an $\sim 7$ \% level of absorption during transit. This absorption can be caused by an opaque disk with a size of $\sim1.7$\,\rp\  that is smaller than the Roche lobe size ($\sim 2.84$\,\rp\, along the planet-star line and $\sim 2.06$\,\rp\, across) of the exoplanet. To properly constrain the atmospheric composition, sophisticated models that include photochemistry and radial transport in the atmosphere of HD\,189733b are needed. In addition, a careful analysis of the transit absorption requires a parametric study that is beyond the scope of this paper. However, using simplified modeling that accounts for the hydrodynamical transport and ionization of the main species in the extended atmosphere is enough for our purpose to achieve a better insight into the O\,I density column required to fit the observed transit drop-off of $\sim 7\%$. Recent photochemistry and hydrodynamic models show that the base of the planetary wind is close to the pressure level of $\sim 0.1-1\mu$bar at which H$_2$\, is dissociated \citep{kos10,mos11}. In the following, we assume that the atmosphere is spherically symmetric and that O\,I is dragged out to high altitudes by H\,I \citep{mun07,kos12a}. This allows us to build the atomic oxygen distribution versus altitude from the H\,I distribution embedded between the bottom of the planetary wind region and the edge of the Roche lobe, assuming a hydrostatic regime for the atmosphere \citep{guo11}. For reference, the H\,I and temperature distributions used here are based on a multi-fluid model of atomic-proton mixture, yet it does not include photochemistry nor heavy atoms \citep{guo11}. 
In a first step, we assume that only thermal broadening affects the atmospheric absorption. Starting from the solar abundance distribution provided by \cite{guo11}, we can estimate an $\sim 3.5\%$ attenuation during transit that corresponds to O\,I density column of $\sim 8.3\times 10^{15}\,{\rm cm^{-2}} $, and a volume density $n_{OI}\sim 6.8\times 10^7\,{\rm cm^{-3}}$ at the base of the model near r=\rp. Scaling the initial atmospheric model to a 40x solar-abundance level produces an attenuation of $\sim 5.3\%$ during transit. The corresponding O\,I density column is $\sim 3.3\times 10^{17}\,{\rm cm^{-2}} $ produced with a volume density $n_{OI}\sim 2.7\times 10^9\,{\rm cm^{-3}}$ at $r=$\,\rp. These high densities are probably difficult to justify because that would increase the atmospheric mean mass, descrease the corresponding scale height, and consequently would require unrealistically higher stellar EUV input to maintain the extended atmosphere \citep{kos10,kos12b}. 

In a second step, we include  super-thermal line broadening that is produced by a layer of hot O\,I atoms confined on top of the atmosphere \citep{ben10}. As described in detail in \cite{ben10}, both the bottom position and the effective temperature of the layer are free parameters. Our intent is not to conduct a parametric study, but rather to investigate how super-thermal populations may modify the result obtained above for only thermal brodening. Assuming solar abundances, a hot O\,I layer above $\sim 1.1$\rp\, with an effective temperature of $T_{OI}\sim 8.4\times 10^4$\,K produces a transit absorption of $\sim 4.6\%$ that is $\sim 1 \sigma$ below the nominal value. Increasing the solar abundance by x5 solar for the same hot O\,I layer produces the observed $\sim 6.4\%$ transit drop-off. If we place the bottom of the hot layer near $\sim 1.5$\,\rp, the same 5x  super-solar model with the same effective temperature of $\sim 8.4\times 10^4$\,K leads to a transit attenuation of $\sim 5.2\%$. Other solutions exist and may require a full sensitivity study.

It is important to stress that our results are not senstitive to the O\,I line selected for the atmospheric analysis. In agreement with recent results on HD\,209458b, super-solar abundances with thermal broadening and  super-thermal broadening probably affect the FUV transit of hot exoplanets \citep{ben10,kos12b}. Sophisticated models that combine photochemistry and hydrodynamic transport in the atmosphere of HD\,189733b are needed in the future to test whether solar or super-solar abundances in the exoplanet's bulk atmosphere may produce enough thermal and  super-thermal O\,I atoms to fit the densities derived here  \citep{mun07,kos12a}. In addition to a sensitivity study that includes most important line brodenings, future observation of high-resolution line profiles of key FUV lines must be obtained with high {\it S/N} to separate the different effects. 

In contrast to the neutral O\,I component for which a classical transit light curve is obtained, the peculiar signature detected in the  early ingress of the C\,II lines seems to indicate that the gas topology around the very XUV-hot Jupiter HD\,189733b cannot be sketched by the Roche lobe volume simply filled by the magnetized plasma. Future observations are much needed to confirm the veracity of the C\,II light curve, yet we can already explore whether such a feature could be predicted by model simulations using realistic parameters for the \hdn\, system.

\subsection{Model simulation of \hdn b magnetosphere}

An early-ingress absorption feature was already reported in the near-UV for the very-hot-Jupiter WASP-12b \citep{fos10,has12}. The first explanation proposed thus far is that the extra absorption is produced by the denser and hotter gas of the bow-shock upstream of the explanet magnetosphere \citep{fos10,vid10,kho12}. To study the stellar wind interaction with the exoplanet, a few numerical simulations have been made using MHD and  hybrid 3D codes \citep{pre07,joh11,coh11a}. While large-scale simulations that include the star-planet system are very useful for obtaining a global view of the system, they do not give that much detail on the planetary magnetosphere itself or on its opacity imprint on the transit light curves. In addition, the intrinsic rotation of the magnetized planet and the formation of the magnetospheric current sheet are not properly incorporated in those 3D numerical simulations \citep[for more details, see][]{kho12}. Therefore, up to now, only empirical models have been used to derive the bow-shock position and predict its opacity effect on the transit light curves \citep{kho12,lla11,tra11}. 

Another explanation of the early-ingress absorption relies on gas-dynamics simulations of the stellar wind interaction with the exoplanet's atmosphere where no magnetic fields are included. These simulations predict a large-scale asymmetric gas distribution around an unmagnetized exoplanet that is supposed to produce a distorted light curve during transit \citep{bis13}. While the stellar wind and planetary magnetic fields are difficult to estimate in general, the gas-dynamics models may produce very useful scenarios of stellar interaction for very weakly magnetized exoplanets, particularly if realistic transit light curves are provided to support the claimed predictions.

In the following, we assume that the early absorption feature observed for \hdn b could be caused by its magnetosphere. To test this scenario, we need to estimate the stellar wind conditions at the planet's orbit at $0.031$ AU as well as the  spatial orientation of the nose of the magnetosphere relative to the planet-star line, taking into account the orbital velocity of the planet \citep{vid10}. \hdn\, is a main-sequence K star for which a coronal stellar wind is expected as revealed by the intense X-ray emission observed \citep{for11, lec12}. To model the stellar wind formation, we used in a first step the Parker model which represents a good approximation despite its limitations \citep{par58,coh11b}. X-ray observations show that the \hdn\, corona has 
\begin{table*}
\caption{\hdn\, stellar wind and magnetosphere parameters}
\begin{center}
\begin{tabular}{cccccc}
\hline
Density (${\rm cm}^{-3}$) & Temperature (K) & Speed (km/s) & B$_{IMF}$ (mG)$^{(a)}$  & Magnetopause nose $\theta$ angle$^{(b)}$  \\
 $<5\times10^7$ & $<4\times10^6 $ & $\sim 200-900$ &  4-23 & $\sim (10-30)^{\circ}$ \\
\hline
\end{tabular} 
\end{center}

\tablefoot{All quantities are estimated at the exoplanet's orbital distance of $0.031$\,AU. The wind parameters are based on the Parker model and a hydrostatic distribution of isothermal gas density. The bottom conditions of the stellar corona are defined from X-rays observations (see text). (a) The interplanetary magnetic field is based on polarimetry observation of HD\,189733 \citep{far10}. (b) The $\theta$ angle is derived using an orbital velocity of $\sim 152.5$\,km/s and a stellar corotation speed of $\sim 28.5$\,km/s of the stellar plasma and magnetic field at the distance of the planet \citep{vid11}. A  spin period of 11.9 days is assumed for the star \citep{hen08}. \label{tbl-1}}
\end{table*}
a temperature in the range of a few $ 10^5-10^6 $\,K and a density up to $\sim 10^{10}\, {\rm cm^{-3}}$ \citep{for11}. We accordingly assumed these properties as boundary conditions for the base of the stellar wind in the Parker model. Combining the isothermal Parker model with a hydrostatic distribution of the gas provides the stellar wind parameters (speed, density, and temperature) at the planet's orbit as summarized in Table 4. For the interplanetary magnetic field (IMF) at the orbit of the exoplanet, we assumed a field strength in the range  $\sim 4-23$\,mG reported for the average field from polarimetry observations of \hdn\, \citep{far10}. At the level of our exploratory work, we can free ourselves from accuracy concerns of the Parker model and accept any solution for the stellar wind properties within the range displayed in Table 4. Taking into account these wind parameters, the orbital motion of the planet, and the stellar corotation effect, we obtain a geometrical orientation of the magnetosphere with a nose deflection angle in the range $10-30^{\circ}$ from the exoplanet-star line \citep{vid11}.  

To quantitatively estimate the opacity impact of the magnetosphere on the stellar flux around transit, we first need to model the HD\,189733b magnetosphere using the stellar parameters in the range described in Table 4. Here, we assume that the planetary magnetic field is a dipole, whose strength is a free parameter in the range $\leq14$G \citep{rei10}. Because our goal is to derive the large-scale structure of the exoplanet's magnetosphere, we used a particle-in-cell (PIC) 3D electromagnetic and relativistic code \citep{bar07,bar11,wod09}, which we describe in the following section.
 
\subsubsection{PIC simulation of magnetosphere and comparison with observations}

We employed a PIC code, originally developed and validated for Earth's magnetosphere \citep{bun93,bar07,bar11}, which we extended to \hdn b using the fact that the expected magnetospheric cavity is relatively small \citep{wod09}. Indeed, assuming a typical magnetic field strength of $\sim7$\,G \citep{rei10} and a stellar wind of $\sim 500$\,km/s, the magnetopause position is estimated to be $\sim 6-8$\,\rp\  comparable to $\sim 10$\,\rp\  on Earth. This allowed us to use a relatively dense grid to describe the whole system without requiring very expensive computer resources. Electrons and ions are represented as macro-particles that contain a large number of real particles. The PIC code solves Maxwell equations on the assumed grid

\begin{equation}
 {\partial { B}\over \partial t} = -\bigtriangledown\times { E},
\end{equation}
\begin{equation}
 \epsilon_o {\partial { E}\over \partial t} = \mu_o^{-1} \bigtriangledown\times { B} -{ J},
\end{equation}
where ${ J}=\Sigma \left( n_i q_i v_i  - n_e q_e v_e\right) $ is the current vector, and follows each macro-particle in the simulation box using the Newton-Lorentz motion equation,
\begin{equation}
 {d(\gamma m{ v})\over dt} = q\left( { E} + { v}\times { B}\right),
\end{equation}
where $\gamma = \sqrt{1\over 1-({v\over c})^2}$ accounts for the relativistic motion of particles. The PIC code units are such that the dielectric constant  $\epsilon_o = 1$ and the magnetic permeability of vacuum $\mu_o = {1\over c^2} $. Because charge is conserved and magnetic fields are divergence-free, both Gauss laws appear as initial conditions. In other terms, since the Gauss laws are fulfilled at t=0, they will remain satisfied at any future time step of the simulation \citep{vil92,bun93}.

To avoid numerical instability,  the Courant condition $c \Delta{t}< {\Delta{r}\over \sqrt{3}} $ should be fulfilled \citep{cou28}; here the speed of light speed was taken to be $c=0.5$, $\Delta{t}$ is the step time, and $\Delta{r}$ is the unform grid size. Plasma instabilities can be efficiently reduced with the strong condition on the plasma frequency $\omega_p.\Delta t \le 0.25$ \citep{tsk07}. In addition, to avoid the classical problems of grid heating and instabilities, we ensured that the Debey length did not reach values below a critical size defined by $\lambda_D \ge { \Delta{r}\over \pi}$ \citep{cai03,tsk07}. To enforce the shielding effect of charges on the Debye volume, we assumed an initial particle density of five pairs per simulation cell \citep{rei02}. In the PIC code, the strength of the magnetic field of the planet is selected to obtain a magnetopause position at the desired value. An initial  estimate was obtained from empirical models of the magnetopause position \citep{vid10,kho12}. Non-reflecting boundary conditions were applied to the fields on all external facets of the box \citep{lin75}. Corotation with the planetary magnetic field was included by adding a corotational electric field as a boundary condition at the surface of the planet. Particles traveling down toward the surface are lost and thus removed from the simulation box. This explains the empty volume obtained around the planet's position (see Fig. 7). 

The simulation box has dimensions of $305\times195\times195$\,$\Delta{r}^3$  along the OX-OY-OZ axes, each pixel  having a uniform size in the range  $\Delta{r} = 0.2-0.5$\,\rp\ depending on the size of the magnetospheric cavity and limitations related to our computer resources. Simulations started with a total number of $5.5\times 10^7$ pairs of macro-particles in the box. The OX axis corresponds to the impinging stellar wind direction relative to the planet and makes the angle $\theta$ with the planet-star line (e.g., Table 4). The OZ axis coincides with both the planet and star spin axes. The XY plane is taken as the equatorial plane of the planet that here concides with the orbital plane. Indeed, the HD\,189733 spin axis is believed to be aligned with orbital plane of the planet \citep{coh11a}. The ion-to-electron mass ratio used in the code 
\begin{figure}
\centering   
 \includegraphics[width=9.cm,height=8.2cm,angle=0]{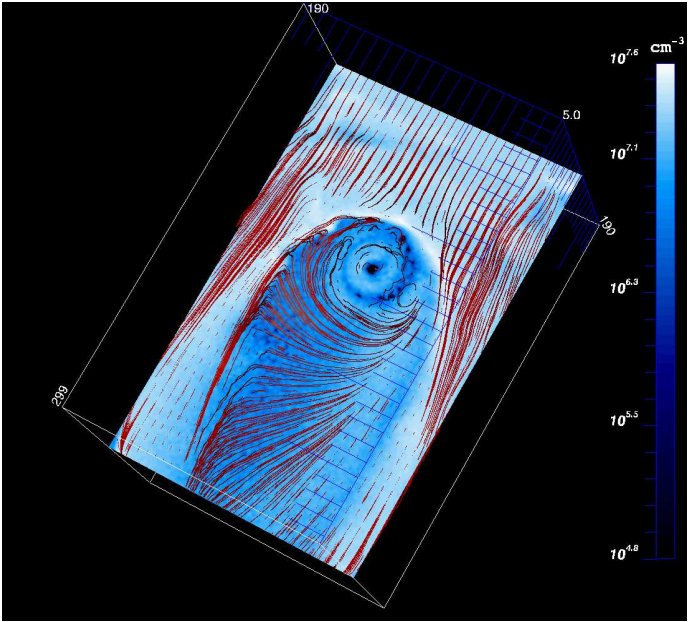}
\caption{3D PIC code simulation of the \hdn b magnetosphere as seen from the {\it bottom} of the simulation box.  The equatorial plane is shown. We can clearly see the exoplanet's bow-shock and atmospheric nebula. Streamlines show the large-scale flow pattern, particulaly the corotation effect of the planet. This simulation was obtained for a stellar wind that has a sonic Mach number $M_s\sim 4.8$, an Alfv\`en Mach number $M_A\sim 5.5$, and a magnetosonic Mach number $M_ms\sim 3.6$. The exoplanet is located at coordinates $x=120\Delta$r, $y=97\Delta$r, and $z=98\Delta$r in the simulation box (reference frame defined by the star-planet line, the north-south magnetic poles, and the dawn-dusk directions). The intrinsic magnetic field of the planet is selected to obtain a stand-off distance of $\sim 16.7$\rp , which allows a good fit to the C\,II light curve (see Fig.8 ). The grid has $\Delta$r units. The star is located toward the top of the page. }
\end{figure}
$mi/me = 32$ is high enough to obtain a good separation between opposite charges and helps complete the simulation in a reasonable amount of steps \citep{gar12}. Finally, the code parameters selected thus far yield an ion skin depth $\lambda_i\sim 12\, \Delta{r}$, a value that ensures that the magnetospheric cavity is properly sketched with the selected grid \citep{omi04,mor12}. We stress here that our goal was not to resolve the detailed kinetic properties of any specific region of the magnetosphere but rather to model its large-scale geometry in a coherent way, which should help test its level of absorption during transit at specific ion resonance lines. 

We modeled the extended atmosphere as a cloud of pairs of macro-ion and macro-electron pairs that are randomly and continuously injected to produce a uniform spherical flow around the planet that is confined in space and that has the same kinetic temperature $\sim 10^4$\,K, density of few $10^7\, {\rm cm^{-3}}$, and radial speed $\sim 5$\,km/s \citep{guo11}. The spatial extent and the total content of the resulting nebula were free parameters that were to be fixed by the transit absorption depth. Ideally, one should include an altitude-dependent ionospheric distribution, but that is beyond the scope of this preliminary study \citep{joh09}. Initial conditions require Maxwell distributions for the velocity field of both macro-ions and macro-electrons at their corresponding temperatures. Stellar wind was injected versus time from the YZ plane located on the top edge of the box in the OX axis (see Figure 7). The planet's dipole magnetic field was switched on smoothly until it reached its maximum value. The system was then left evolving to reach a steady state in which the observable large-scale structures were not changing anymore. Post-processing of simulation results allowed us to derive all 
\begin{figure}
\centering
\hspace*{-0.2in}
 \includegraphics[width=8.5cm, angle=90]{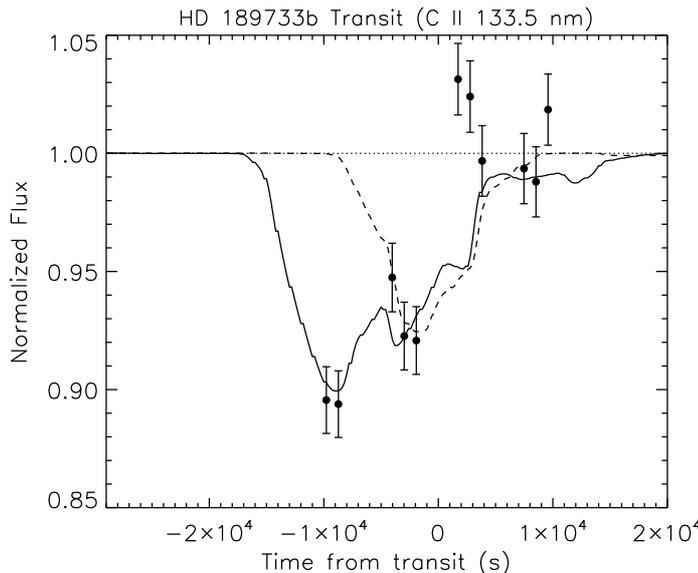}
\caption{3D PIC simulation Fit to \hdn b C\,II 133.5\,nm light curve. The simulations are obtained for the stellar plasma parameters indicated in Table 4. A solar C/H$=2.45\times 10^{-4}$ is assumed \citep{asp05}. (Dashed) Model assuming a magnetic field strength of $7$\,G for the exoplanet and a fast stellar wind (speed $\sim 500$\,km/s, density $\sim 6\times10^6 {\rm cm^{-3}}$, and temperature $\sim 4\times 10^6$\,K). (Solid) Using a magnetic field strength of $\sim 5.3$\,G for \hdn b, and stellar wind speed of $\sim 250$\,km/s, temperature $\sim 1.2\times 10^5$\,K, and density $\sim 6.3\times 10^6\,{\rm cm^{-3}}$ lead to a better fit corresponding to a MP position of $\sim 16.7$\,\rp\, upstream. This solution is not unique (see text).}
\end{figure}
moments of the macro-particles velocity distributions anywhere in the box. Plasma number density, temperature, and bulk velocity distributions were thus calculated, allowing us estimate the gas opacity of the magnetosphere at the desired spectral line and its obscuration of the stellar disk flux as the planet moves along its orbit from ingress to egress. In Figure 7, a typical magnetospheric configuration is shown in the equatorial plane. We can easily distinguish the planetary atmospheric nebula, the plasma equatorial sheet, and the upstream bow shock. Streamlines of the bulk speed in that plane show the corotation effect on the plasma located near the planet. Here, it is important to stress that the equatorial plasma and current sheet (also called magnetodisk) is naturally formed with the PIC code, a consistency that cannot be met in MHD simulations. 

In a first step of the simulations, assuming a surface magnetic field strength of $\sim 7 $\,G \citep{rei10} for the planet and using the fast stellar wind solution as derived from X-ray observations ($\sim 500 $\,km/s ), a magnetospheric configuration with a nose  angle of $ 10-30^{\circ}$ from the star-planet line produces only a poor fit to the C\,II light curve (assuming C/H solar abundance; see Fig. 8). It is important to notice that we do obtain an asymmetric transit light curve and early-ingress absorption, but later than observed. The model misfit tends to show that the magnetopause (MP) position obtained by the model ($\sim 7$\,\rp) is too close to the planet to enable a fit to the observed C\,II early-ingress absorption seen in HST orbit 1 (see Fig. 8).  

To improve our diagnostic, we therefore focused on the magnetopause position that allowed to fit the observed early-ingress absorption. In practice, we varied the ram pressure of the stellar wind until a good fit was obtained for the C\,II light curve, keeping the magnetic field strength of the exoplanet assumed in the PIC code unchanged. As shown in Figure 8, a good fit is obtained with an MP upstream stand-off distance of $\sim 16.7$\,\rp. It is interesting to notice that in addition to the transit signature due to the planetary atmosphere (at $\pm3000$\,s from mid-transit), early absorption by stellar plasma in the bow-shock region clearly appears on ingress (around 10,000\,s before mid-transit), while a weak absorption also appears on late egress ($\sim$4000--10,000\,s after mid-transit), probably from plasma in the magnetotail (see Fig. 8). We emphasize that because the magnetopause is a stagnation region that results from total pressure balance between the star and the planet, other solutions for the planetary magnetic field and stellar wind parameters may lead to the same MP position. Furthermore, because the stellar parameters are not accurately known, the orientation of the magnetosphere is not unique, which leads to a degeneracy in the possible solutions of the projected size of the magnetospheric cavity. 

For example, for a stellar wind speed of $\sim 250$\,km/s, a temperature of $\sim 1.2\times 10^5$\,K, and a density $\sim 6.3\times 10^6\,{\rm cm^{-3}}$ at the planet orbit, the planet's magnetic field should be $\sim 5.3$\,G to fit the MP position at $\sim 16.7$\,\rp\ . Assuming the same stellar wind parameters, but with a speed of $\sim 450$\,km/s, our model simulation requires an exoplanet magnetic field of $\sim 7.2$\,G to obtain a good fit to the C\,II transit light curve. As a final example of a good fit, stellar wind parameters with a speed of $\sim 250$\,km/s, a temperature of $\sim 5.0\times 10^5$\,K, and a density $\sim 1.3\times 10^7\,{\rm cm^{-3}}$ at the planet orbit along with an exoplanet magnetic field of $\sim 12.5$\,G also give a satisfactory fit to the C\,II transit light curve. All these results can be easily understood from the total pressure balance that occurs at the MP. Indeed, if the stellar wind total pressure is changed for the same MP position, the total planetary pressure should be modified on the opposite side to maintain the pressure balance. In that way, an excessively strong planetary magnetic field could be required to fit the MP position at $\sim 16.7$\,\rp\, for some stellar wind parameters. For example, for a stellar wind temperature of $\sim 10^6$\,K, a density of $\sim 1.0\times 10^7\,{\rm cm^{-3}}$, and a speed of $\sim 250$\,km/s, a good fit could be obtained but with a planet magnetic field of $\sim 16.2$\,G. Finally, we emphasize that we could find models that produce a magnetosphere with the required MP position, but that do not fit the C\,II transit light curve. A full sensitivity study versus the stellar wind and planetary atmosphere parameters is needed, but this is beyond the scope of the present study.

In practice, to improve the present diagnostic of a magnetospheric signature during transit, we need to obtain a better coverage of the transit light curve of strong resonance lines of several ionized species, particularly on ingress and egress (see Fig. 8). Additionally, to account for potential orbital phase variations in the stellar wind, we may need models that include the full stellar corona-exoplanet system in the simulation box to better uncover any star-planet electromagnetic coupling \citep{coh11a}. For all these reasons, only a sensitivity study may help separate the different effects and properly constrain the exoplanet magnetic field if a full transit light curve is confirmed with the expected strong early-ingress or weak late-egress absorptions. To that end, recent studies on the solar wind interaction with the magnetized local interstellar medium can provide a good reference for a future approach to uncover the plasma environment and magnetic field of \hdn b \citep{ben12}.

\section{ Summary and conclusions} 

Using archival HST/COS G130M observations, we have positively detected neutral oxygen $\sim7\%$  absorption during the transit of exoplanet \hdn b. We also reported a tentative detection of singly ionized carbon. If sketched as a compact disk, this detection of O\,I reveals a distribution of atoms extending to about $\sim 1.7$\,\rp\, inside a Roche lobe radius of $\sim 2.84$\,\rp. Assuming a mean temperature of $\sim (8-12)\times10^3$\,K for the upper atmosphere of the exoplanet, we used the hydrodynamic  model of \cite{guo11} to derive that for solar abundances with an O\,I density column of $\sim 8\times 10^{15}\,{\rm cm^{-2}}$, an attenuation of $\sim 3.5\%$ is obtained for the O\,I 130.6\,nm line during transit, at least $\sim 2.5\sigma$ below the observed absorption. Invoking up to 40 times solar abundances with an O\,I density column of $\sim 3.3\times 10^{17}\,{\rm cm^{-2}}$, produces an attenuation of $\sim 5.3\%$ during transit that is closer to observations. These very high density columns of heavy atoms are difficult to justify because they would require substantial additional energy in the system to compensate for the reduced scale height. In contrast, we show that including a  super-thermal O\,I layer embedded on top of the atmosphere reduces the need for highly super-solar abundances. For example, with twice solar abundaces and a hot O\,I layer confined above $\sim 1.5$\,\rp\, with an effective temperature of $T_{OI}\sim 8.4\times 10^4$\,K, a transit absorption of $\sim 4.3\%$ is produced that is only about $\sim 1 \sigma$ below the observed value. Our preliminary results tend to confirm the general conclusion that similar to HD\,209458b, super-solar abundances with thermal broadening as well as  super-thermal broadening probably affect the FUV transit of hot exoplanets \citep{ben10,kos12b}. High-resolution line profiles observations of key FUV lines must be obtained with high {\it S/N} to separate the different effects. The O\,I detection reported here opens up new perspectives to use the high sensitivity of the HST/COS instrument to detect oxygen in extrasolar systems in which the stellar O\,I lines are thin and bright enough compared to the broad geocoronal line that results from the extended sky background filling the 2.5" aperture.

Furthermore, we found a peculiar signature in the C\,II 133.5\,nm transit absorption by \hdn b that  shows different properties and a dependence on orbital phase, which appears to be as an early-ingress absorption. Combined multi-spectral analysis and a comparison of line profiles support the reality of the feature. However, because the star is relatively active (e.g., Fig. 6), both the transit and the early-ingress detections require confirmation in the future to determine any stellar variability that cannot be controlled from the present analysis of only one single transit observation.

 Assuming the extra absorption is real, we used the Parker model for the stellar wind and a PIC code simulation of the interaction of the magnetized and extended thermosphere of \hdn b with the impinging plasma wind. First, our analysis showed that a magnetosphere forms with a bow shock and a nose that is oriented almost $\sim 10-30^{\circ}$ from the star-planet line for most cases of stellar wind parameters. Second, the simulations revealed that a magnetopause stand-off distance of $\sim 16.7$\,\rp\ is required to obtain a satisfactory fit to the transit light curve observed for C\,II (e.g. Fig. 8). However, our preliminary assessment of the PIC simulation results showed that many solutions exist for the stellar wind parameters and the magnetic field of the exoplanet that fit the C\,II transit curve. Several reasons could explain the origin of this degeneracy, namely the limited coverage of the transit event by the HST/COS observations, the signal variability (stellar or instrumental), and finally the nature of magnetospheric cavity. Indeed, because the magnetopause is a stagnation region where a total pressure balance occurs between stellar and planetary plasmas  and fields, no unique solution exists for either the stellar wind parameters or the magnetic field strength of the exoplanet that provide a prescribed magnetopause stand-off distance. For example, we are able to find a good fit for the C\,II transit light curve with a speed of $250$\,km/s, a temperature of $\sim 1.2\times 10^5$\,K, and a density $\sim 6.3\times 10^6\,{\rm cm^{-3}}$ at the planet orbit for the stellar wind, and a magnetic field strength of $\sim 5.3$\,G  for the exoplanet magnetic field. However, other solutions exist (see previous section). 

In all cases, whether the C\,II peculiar feature is real or not, our PIC simulations showed that the C\,II spatial  distribution that contributes to the observed transit absorption is not solely related to the planet's extended atmosphere but rather is dependent on the plasma configuration that results from the complex interaction between the stellar wind, the exoplanet atmosphere, and the local planetary and stellar magnetic fields. New HST FUV observations are urgently needed to confirm our results and extend the coverage of the early-ingress and late-egress behavior in the C\,II 133.5 nm line absorption. Our quick comparison of PIC simulations of \hdn b's magnetosphere with transit observations showed that fundamental properties of the plasma configuration around the exoplanet may be accurately constrained through the specific shape of the light curve versus wavelength bands either from a single line or using multiple lines.

\begin{acknowledgements}

LBJ acknowledges support from CNES, Universit\'e Pierre et Marie Curie (UPMC) and the Centre National de la Recherche Scientifique (CNRS) in France. GEB acknowledges support from STScI through grants HST-GO-11673.01-A and HST-AR-11303.01-A. The authors thank J. Sanz-Forcada and O. Cohen for clarifications on stellar coronal parameters and the solar and stellar winds. They also thank the referee for several suggestions that helped improve the manuscript. LBJ acknowledges helpful discussions within the ISSI and Europlanet teams working on "Characterizing stellar and exoplanetary environments". This work is based on observations with the NASA/ESA Hubble Space Telescope, obtained at the Space Telescope Science Institute, which is operated by AURA, Inc.
\end{acknowledgements}

\end{document}